# Non-Speckle-based DVC for Measuring Large Deformations in Homogeneous Solids using Laboratory X-ray CT

Zifan Wang[a,*], Akshay Joshi[a,*], Angkur Jyoti Dipanka Shaikeea[a], Vikram S. Deshpande[a]

[a] *Department of Engineering, University of Cambridge, Cambridge CB2 1PZ, U.K.*

## Abstract

X-ray computed tomography (XCT) has become a reliable metrology tool for measuring internal flaws and other microstructural features in engineering materials. However, tracking of material points to measure three-dimensional (3D) deformations has hitherto relied on either artificially adding tracer particles (speckles) or exploiting inherent microstructural features such as inclusions. This has greatly limited the spatial resolution and magnitude of the deformation measurements. Here we report a novel Flux Enhanced Tomography for Correlation (FETC) technique that leverages the inherent inhomogeneities within nominally homogeneous engineering polymers to track 3D material point displacements without recourse to artificial speckles or microstructural features such as inclusions. The FETC is then combined with a Eulerian/Lagrangian transformation in a multi-step Digital Volume Correlation (DVC) methodology to measure all nine components of the deformation gradient within the volume of complex specimens undergoing extreme deformations. FETC is a powerful technique that greatly expands the capabilities of laboratory-based XCT to provide amongst other things the inputs required for data-driven constitutive modelling approaches.

***Keywords:*** *Digital Volume Correlation, X-ray CT, in situ experiments, 3D Strain Measurements*

*Both authors contributed equally

**Introduction**

Over the past decade, there has been a significant increase in the development and use of data-driven techniques to rapidly design[1–4] and characterize[5–11] new materials and structures. This is largely due to the development of powerful computing resources and novel machine learning techniques. These data-driven techniques essentially provide a computationally inexpensive surrogate of material behavior to aid inverse design. They can be broadly classified as those that bypass[5,12] or surrogate[2,7,13–16] material constitutive laws, or those that discover them[8–11]. Most of these techniques train their predictors on stress-strain data which are either artificially generated using regular or multiscale Finite Element (FE) simulations[17]. However, a major obstacle in the practical deployment of most of the data-driven techniques is the lack of availability of high-fidelity experimental stress and strain tensor data. Although some data-driven techniques[8–11] rely exclusively on experimentally obtainable boundary force and displacement field data, they are yet to be validated in generic three-dimensional contexts. Furthermore, there seems to be a paucity of experimental techniques that can provide a reliable 3D volumetrically resolved displacement field and subsequent strain fields for generic engineering materials.

Obtaining volumetrically resolved stress and strain data is challenging since volumetrically resolved stress distribution can only be experimentally obtained for limited configurations and materials. The lack of volumetrically resolved stress tensor data can be compensated for to a certain extent by using the boundary force data and by leveraging the conservation of linear momentum[8,18]. To obtain volumetrically resolved strain distribution using Digital Volume Correlation (DVC) and reliable tomography (e.g., confocal, MRI), some heterogeneities have to be tracked in the sample. This can be done either by artificially adding tracer particles or by leveraging inherent microstructural heterogeneities in the material. Meanwhile, the resolution of displacement measured using heterogeneities is limited, and current tomography techniques are unable to register sufficient micro-heterogeneities in nominally homogeneous specimens to enable accurate DVC.

This study aims to demonstrate a novel approach to quantify extreme deformations on homogeneous solids (a range of polymeric solids with physical densities between $1000 - 1500$ kgm$^{-3}$) using a lab-based X-ray system, without the need for tracer particles/speckles. Previous attempts[19] have either relied on inherent textures in microstructures as available in bones and ceramics or have added small volumes (typically 5-15 % v/v) of tracers like copper powder to track displacements within the 3D volume[20]. We introduce a method that increases the sensitivity of X-ray attenuation to minor fluctuations in physical density/textures (solid at the resolution of the 3D scan) in the homogeneous solid. This is achieved primarily by increasing the flux of the X-ray source, and hence we refer to it as Field Enhanced Tomography for Correlation (FETC). The grey scale variations registered from the materials inherent inhomogeneities are leveraged as speckles (in DVC analysis) that can be tracked between successive scans, enabling accurate displacement measurements at material points. The FETC is then used in conjunction with a Eulerian/Lagrangian transformation in a multi-step DVC methodology to accurately measure all nine components of the deformation gradient within complex specimens that undergo extreme deformations.

We demonstrate FETC coupled DVC for hyperelastic materials in different specimen geometries and loading conditions. We used the commercial software VGStudioMax 2022.4 to track the displacement of every material point using the global DVC approach[21–23]. The material points were recognized using their respective grayscale patterns in the first frame, and DVC was performed between consecutive tomography frames, i.e., with the previous tomography as a reference. This enabled us to obtain frame-wise deformation gradients at every material point. We obtained deformation gradients by performing finite element differentiation on de-noised displacement data. In the following sections, we first explain the novel FETC approach and then elaborate on the multi-step DVC for extreme deformations. Results on three different scenarios are discussed followed by demonstration of the FETC technique to other engineering polymers to emphasis the generality of the approach.

# Results and discussion

## 2.1 DVC in a homogeneous solid without tracer particles

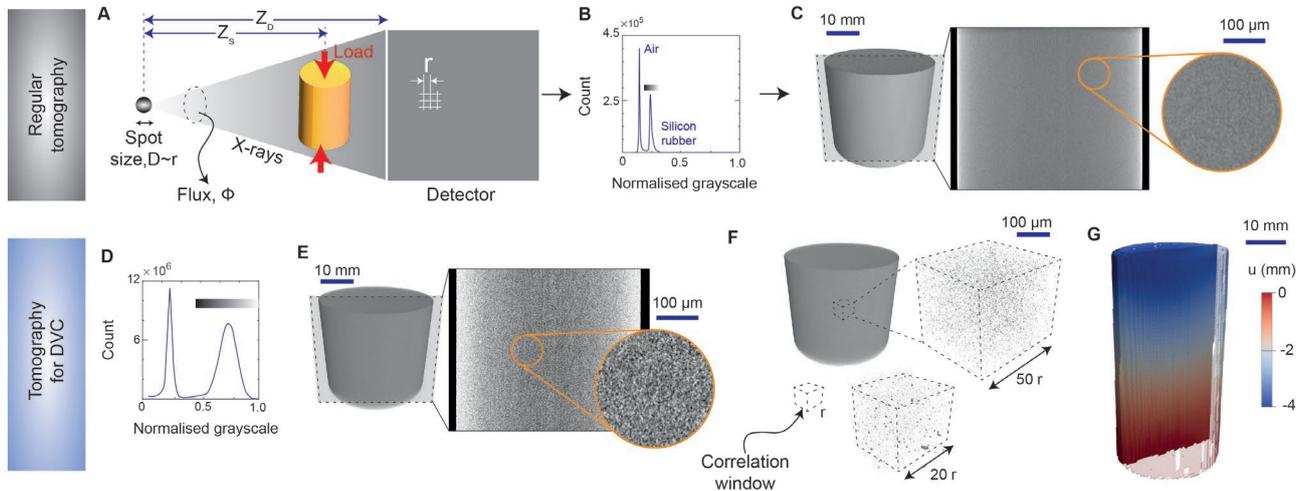

**Figure 1:** Overview of the Flux Enhanced Tomography for correlation (FETC). (a)-(c) Regular tomography wherein the X-ray source parameters are adjusted to give sharp transmission peak for material and air to get sharp boundaries between the material and air. This results in a featureless tomograph of material with no speckles for conducting Digital Volume Correlation (DVC). (d)-(g) FETC wherein we spread the transmission peak to pick up small property variations in the nominally homogeneous material. These variations provide a natural 3D speckle pattern for conduction DVC.

In X-ray imaging, an object placed between an X-ray source and detector produces a greyscale histogram that measures the attenuated X-rays. Here we use a cylindrical polymer specimen (casted silicone rubber, MBFibreglass Polycraft GP3481-F RTV Condensation Cure, handing mixing, ratio 20:1) of diameter ~20 mm and height ~30 mm and perform in-situ XCT with uniaxial compression (Fig. 1a). The histogram (Fig. 1b) is a signature on the silicon rubber's physical density and atomic weight. Ideally a continuum homogeneous solid would show a singular grey value, but in practice a histogram of grey values is recorded. To obtain an artifact-free tomography image, it's essential to keep the power to a minimum (limited by the resolution) and separate the material peak from interfaces such as air. In Fig. 1c, a clean tomograph is shown, with the air peak separated from that of material and the silicon rubber appearing as a distribution of grey values (Fig. 1b). A slice taken along the vertical mid-section of the 3D volume is plotted as a linear gradient in grey values (Fig. 1c) across the silicon rubber peak, with all regions to the left of the material peak blacked out (Fig. 1b). Magnified images emphasize the smooth grayscale variation within the solid. However, in this study, we will alter this general protocol of X-ray tomography by increasing flux (current) and add filters (Copper and Aluminum) to the source to cut off low-energy X-rays, resulting in a much wider spread of the histogram across the specimen (Fig. 1d). We refer to this method of scan acquisition as Flux Enhanced Tomography for Correlation (FETC). The wide range of grey values represents minor fluctuations in texture/physical density variations originated from the casting process. These fluctuations are visualized as textures (Fig. 1e) in the same midsection as Fig. 1c but with the revised scan protocol (FETC). The DVC algorithm[21–23] available in the commercial software VGStudioMax 2022.4 is used to navigate through the various correlation window sizes (Fig. 1f). Iterated over 500 attempts with a window size of ~600 microns give the best correlation coefficient for the DVC analysis. Results for displacement field from the DVC analysis between two stages (uniaxial compression between uncompressed and $u = 1\ mm$) is shown in Fig. 1g.

*2.2 Multi-step tracking of material points for large deformations*

Despite the fact that DVC has been implemented in quantifying the displacement field during *in situ* mechanical loading, the extents of deformations in up-to-date studies are in the range of a few microns[24] up to a few millimeters[25]. This is mainly due to the limitation of maximal displacement that can be quantified in a solitary DVC analysis. Namely, between the initial state and loaded state, the microstructural features undergoing large deformation may not be correctly correlated by DVC algorithms. Depending on the microstructure, materials have different upper limit for a one-step DVC analysis, but the certain thing is that for superelastic material, hyperelastic material, origami metamaterials etc. which exhibit large deformation, one-step DVC fails to quantify the displacement field in such cases. Thereby an effective multi-step approach to quantify the displacement field in a hyperelastic silicon rubber subjected to large deformation under uniaxial compression is demonstrated in Fig. 2.

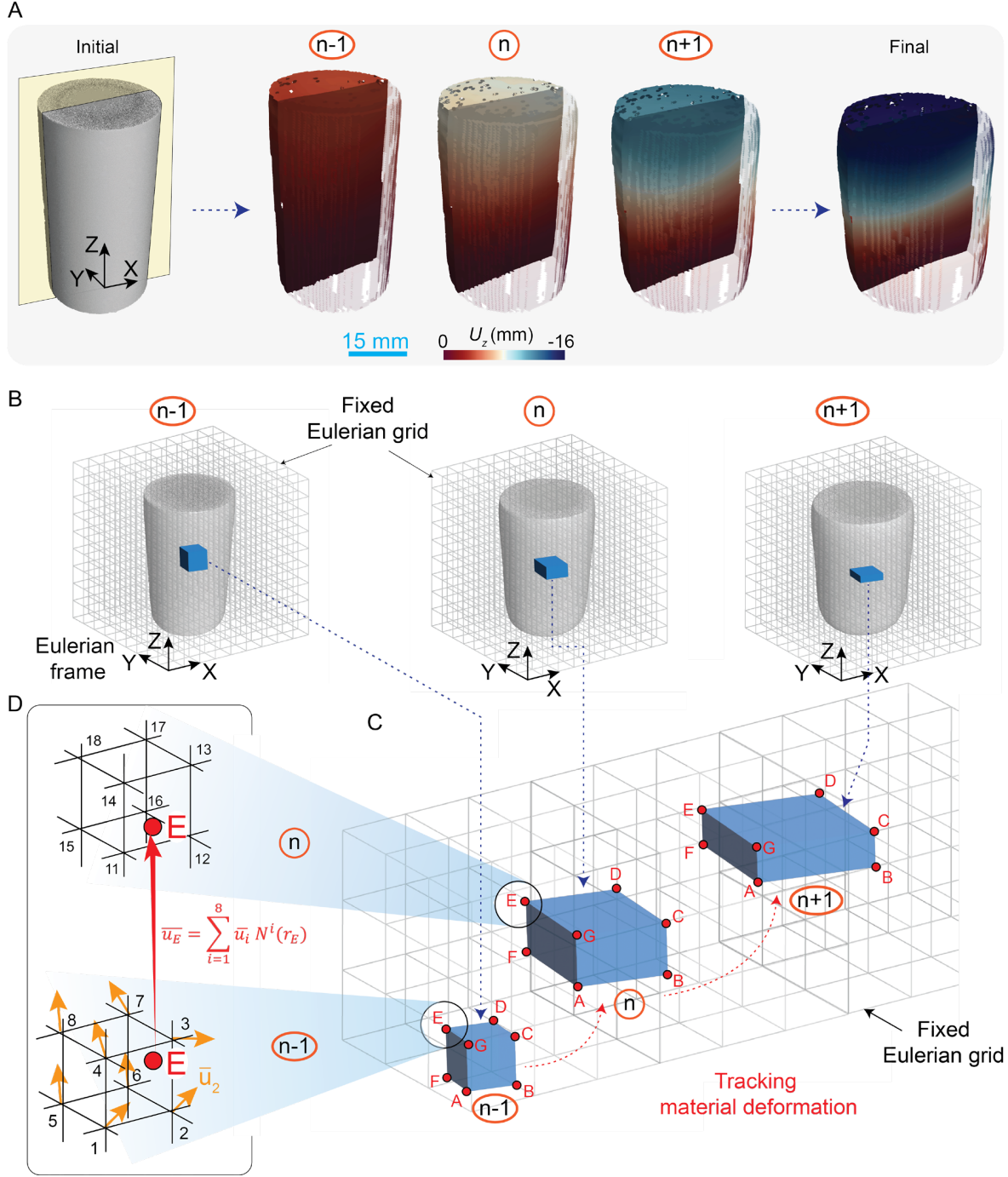

**Figure 2.** Multi-step quantification of 3D displacement field based on DVC analysis. (a) Uniaxial compression of a solid cylinder made of hyperelastic silicon rubber. Four incremental loading steps were conducted. Each loading step has 4mm displacement in Z axis direction, giving a total displacement of 16mm with $U_z$ denoting the displacement in Z axis (vertical direction). (b,c,d) Schematic of the novel Eulerian/Lagrangian transformation algorithm for multi-step quantification of 3D displacement field in the material. Note that the Eulerian frame is fixed in laboratory frame in which the X-ray detector is located.

The solid silicon rubber was loaded in a multi-step manner, as shown in Fig. 2a. The bottom of the sample sits on a fixed platen, while the top is in contact with a platen that can move in Z axis direction in order to apply uniaxial compression. Each loading step has 4mm displacement on the top, hence 16mm in total for four loading steps in the final state. The 4mm displacement was determined to be the maximal amount of deformation that can be accurately quantified by DVC analysis. DVC analysis was performed between each intermediate step, namely, initial and $(n-1)$, $(n-1)$ and $(n)$, $(n)$ and $(n+1)$, so on and so forth. A novel multi-step approach has been developed to track material deformation based on the results of a series of independent DVC analyses. The total displacement field of each step is also given. It can be seen that the multi-step approach is able to quantify the incremental displacement field throughout the loading process, as the top of the cylindrical sample exhibits 16mm displacement downwards in the final state, which well matches the micrometer displacement reading from the mechanical loading rig.

The principle of the multi-step DVC approach is demonstrated in Figs. 2b-d. In Fig. 2b, 3D structure of the sample in each loading step has been reconstructed from the XCT scan in the fixed Eulerian coordinates. The displacement field of each loading step can be obtained from DVC analysis, e.g., the displacement field of $(n-1)$ step comes from the DVC analysis between $(n-1)$ and $(n)$. However, it should be noted that DVC analysis outputs the displacement field in the form of the three orthogonal displacements ($u_x$, $u_y$, $u_z$) at individual points in the 3D grid which is aligned with the fixed Eulerian coordinates. This means that the deformed 3D grid points in step $(n-1)$ are not aligned with those of step $(n)$ where the new displacement field is output from DVC analysis. Consequently, the displacement field of the material cannot be tracked throughout multiple steps.

To illustrate the multi-step DVC approach, the cubic volume of material is taken as an example in step $(n-1)$. In step $(n)$ and step $(n+1)$, the cubic volume changes its shape into a cuboid due to the compression. This shape change can be reflected from the motion of vertexes A-G, as shown in the magnified view in Fig. 2c. Taking vertex E as an example, in step $(n-1)$, its displacement $u_E$ is expressed as a function of its surrounding grid points 1-8:

$$\overline{u_E} = \sum_{i=1}^{8} \overline{u_i} \cdot N^i(r_E) \tag{1}$$

where $u_i$ is the displacement of the grid point obtained from DVC analysis between step $(n-1)$ and step $(n)$, $N^i$ is the linear shape function of the grid point, and $r_E$ is the spatial position of vertex E relative to the grid points 1-8. Note that the grid points are always aligned with the fixed Eulerian coordinates. Hence, the displacement of vertex E is known and its motion being tracked, resulting in a new position in step $(n)$, as illustrated in Fig. 2d. From DVC analysis between step $(n)$ and step $(n+1)$, the displacement of the grid points 11-18 can be obtained, which forms the grid where the vertex E sits. This tracking algorithm is repeated so that new position of vertex E is updated. Through this procedure, the material deformation can be tracked precisely, regardless of the number of steps and the amount of total displacement between the initial and final state. Our multi-step DVC approach overcomes the barrier in the existing DVC techniques that large deformation cannot be effectively quantified[26]. This opens up the possibility to study the deformation mechanisms in a wide range of hyper-deforming materials and structures, including soft tissues, metamaterials, superelastic materials, and so on.

### *2.3 3-point bend – a complex quasi-2D deformation*

FETC coupled with multi-step DVC tracking of material points measures large deformation field of nominally homogenous material subjected to complex loading. Here, we exemplify this approach with 3-point bend test (Fig. 3a) of the hyperelastic silicone rubber. Each loading step has 4 mm displacement in Z axis direction on the top loading tip, hence three loading steps yield 12 mm displacement in total. The material shape and $u_z$ field in every one of the three loading steps is shown in Fig. 3b. Besides the uniaxial compression of solid cylinder in section 2.2, the multi-step DVC methodology is able to capture the complex shape of material during large deformation, meanwhile, the displacement field can also be quantified precisely, that the beam center exhibits $u_z$ = -12mm, and the two ends exhibit positive value due to the constraint of the two static tips. Based on the displacement field, all the nine components of deformation gradient field $F$ can be calculated, and the most typical component $F_{zz}$ is shown in Fig. 3c at each loading step. $F_{zz}$ is the derivative of displacement field in the space along Z-axis, thus reflecting the strain value in Z axis direction. For principal component $F_{xx}$, $F_{yy}$, and $F_{zz}$, a value above 1 indicates tension, whilst below 1 indicates compression. In loading step 1 in Fig. 3c, it is noteworthy that the area below the top loading tip exhibits compression, but gradually turns into tension in loading step 2, and further intensifies in step 3.

A small fluctuation in the displacement field can cause considerable disturbance in the $F_{zz}$ field due to the differentiation. Therefore, the smooth evolution of $F_{zz}$ demonstrates both the continuity and accuracy of displacement field quantified via the multi-step DVC methodology. Following the approach, detailed analysis of 3D displacements and deformation gradient field is displayed in Fig. 4. The results are calculated accumulatively through multiple loading steps between the initial state (Fig. 4a) and the final deformed state (Fig. 4b), and are plotted on the undeformed state.

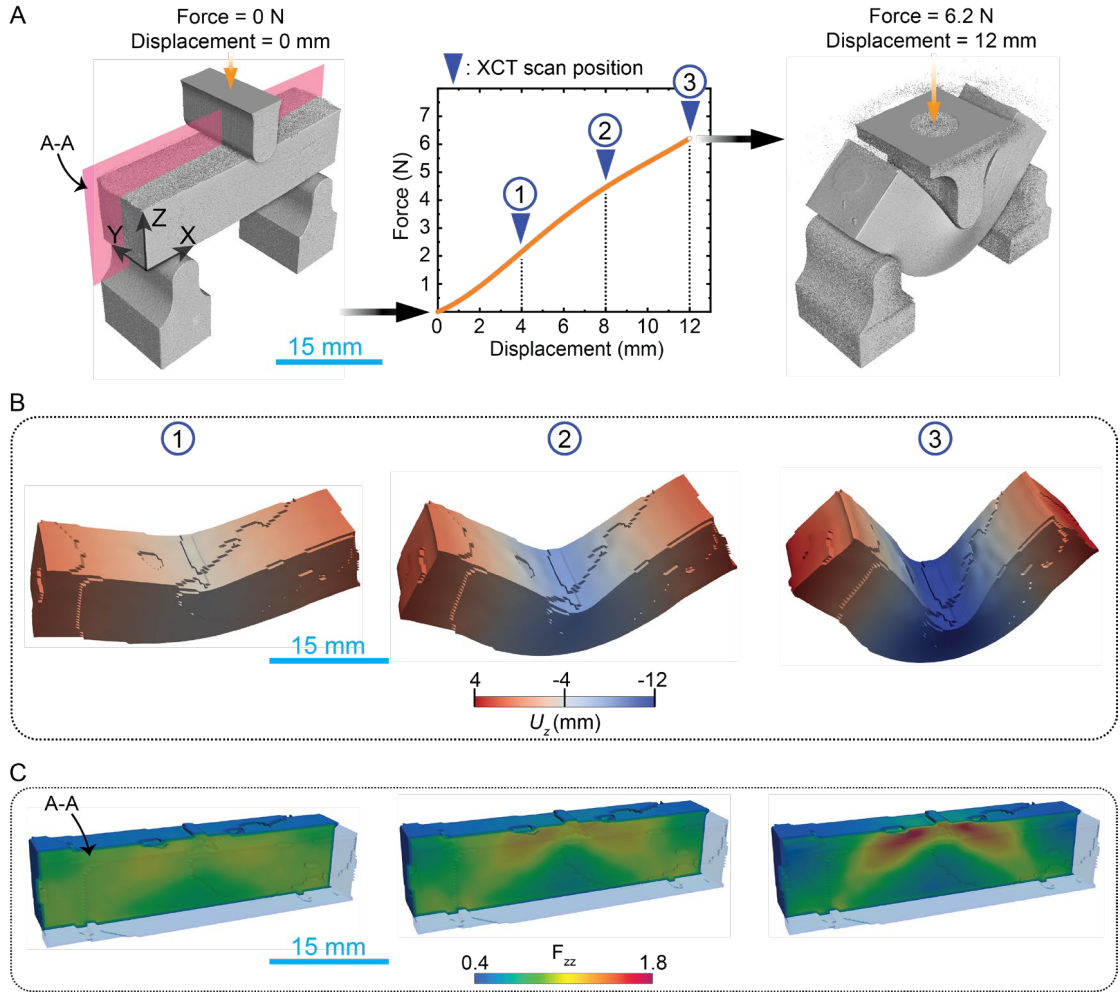

**Figure 3.** Multi-step DVC on 3-point bending test of silicone rubber cuboid beam. (a) XCT reconstruction of the loading setup in the initial (displacement = 0 mm) and final state (displacement = 12 mm). The measured bending force vs. displacement history is also shown. Sample was held in displacement control mode at each XCT scan position. (b) The DVC inferred $U_z$ field of the cuboid beam at each loading step. (c) The spatial distribution of the deformation gradient $F_{zz}$ on the central clipping plane marked in (a) at state 3 (i.e., the final deformed state).

Selected line plots are plotted in Fig. 4c. Along Line 1, it shows $u_z \approx -12$mm on the top surface of the beam, which coincides well with the total amount of displacement from the loading rig micrometer, and gradually reduces to -10mm at the bottom, indicating a compressive deformation in Z axis direction under the top loading tip. $u_z$ along Line 2 indicates the V shape of the beam in the final deformed state. $F_{zx}$ represents the derivative of variation of $u_z$ in X axis direction, therefore for a beam deformed to a V shape in the present study, value of $F_{zx}$ should be positive in one half, negative in the other, and be continuous at the center, which exactly matches with the experimental observation. Due to the significant rigid body rotation in the material upon deformation, the Green-Lagrangian strain $E$, which is commonly used for evaluating large deformation of elastomers, has also been derived and given. The experiment results in these line plots are displayed with the corresponding FE simulation results, however, it should be pointed out that FE simulation results cannot be used as a way of validating experiment result, as the constitutive law is not well established and certain assumptions have been made in terms of incompressibility of the material, friction coefficient between the material and the loading tips, and geometric symmetry of the setup. Instead, it can only be regarded as an elementary check of the variation tendency of the experiment results.

Figure 4d shows the 3D full field of $F$ and $E$ components via various clipping planes. For all the components, only slight variation in Y axis direction can be observed from plane B-B, C-C and D-D. The last row of Fig. 4d presents the $E$ components in comparison to their $F$ counterparts in the first row. Basically, similar strain field can be seen between the same $F$ and $E$ components, whilst E components better represent the strain field in the material as $F$ contains rigid body rotation. Furthermore, the determinant of $F$, also known as "det F", implies volumetric compression for value less than 1, and volumetric expansion for value larger than 1. 3D field of det F indicates significant compression and expansion in the volume throughout the whole material upon deformation, thereby confirming the compressibility of the material, and consequently, proving that any possible FE simulation results are too ideal, and that a more accurate constitutive law for this material is needed.

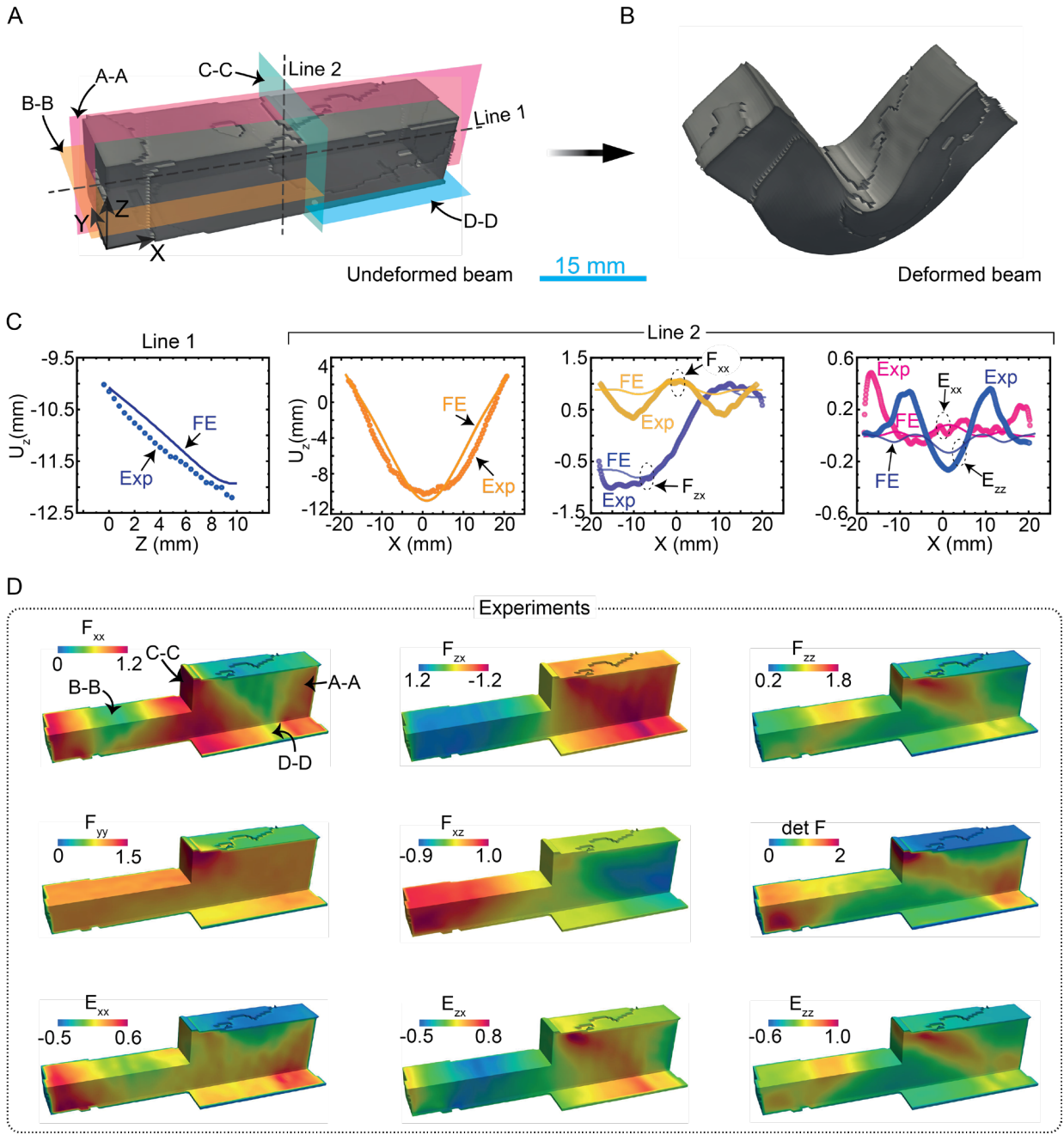

**Figure 4.** The measured 3D deformation field within the 3-point bend specimen. (a) Initial shape on which we indicate planes and lines over which we shall show the measured deformation fields. (b) XCT reconstruction of the final deformed shape of the beam for which all results in (c) and (d) are shown. (c) Variations of the displacement $U_z$ and selected components of the deformation gradient $F_{ij}$ and Green-Lagrange strain tensor $E_{ij}$ along Lines 1 and 2 shown in (a). The corresponding FE predictions using a Neo-Hookean material model are also indicated. (d) Illustrations of the 3D deformation field shown in planes indicated in (a). In addition of selected components of $F_{ij}$ and $E_{ij}$ we also indicate the volumetric deformation via the determinant of $F_{ij}$ and labelled "det F".

*2.4 Compression on cylinder with cubic inclusion – an arbitrary 3D deformation*

Advanced from the complex quasi-2D deformation study in the previous section 2.3, an arbitrary 3D deformation field quantified by the multi-step DVC approach is demonstrated in a hyperelastic silicone rubber cylinder casted with a solid acrylic cubic inclusion in the middle. As shown in Fig. 5a, a four-step compressive loading was conducted, giving a total of 16mm downward displacement on the top surface. Via the multi-step DVC approach, the displacement field is firstly quantified at each loading step, as shown in Fig. 5b. Contrary to the compression test results in section 2.2, the displacement field is "preserved" around the cubic inclusion due to its rigid body motion. Correspondingly, the $F_{zz}$ field has been derived and shown in Fig. 5c. The whole $F_{zz}$ field is in compression (value less than 1), with the region above and below the cubic inclusion being the most intense. The top and bottom regions adjacent to the platens exhibit lowest amount of compression due to the limitation of compressibility, as the geometric constraint exerted by the high friction between the material and the platen (which can also be observed from the circumferential shape in the final deformed state).

An elaborative display of the arbitrary 3D deformation field is shown in Fig. 6. To demonstrate the complex variation of deformation in 3D, various clipping planes have been made, as illustrated in the $F_{zz}$ field in Fig. 6a. Strong variation of $F_{zz}$ can be observed in all directions due to the symmetries. Those tiny hollow regions are air voids introduced during the manufacturing process, which cause disturbance in the deformation field. This phenomenon has been captured perfectly, demonstrating the preciseness of the multi-step DVC methodology.

Other representative components of deformation gradient tensor *F* and Green-Lagrangian strain tensor *E* are shown in Fig. 6b. The $F_{yy}$ field exhibits tension across the whole domain, while the regions adjacent to the top and bottom rig platens have minimal value because of the high friction between the material and the platens, notably, the same effect can be seen in the regions around the cube inclusion. Unlike the huge difference between *F* and *E* fields in the case of 3-point bending of cuboid in Fig. 4d, which is attributed to the rigid body rotation, the distinction is trivial in this situation. Namely, $F_{zz}$ and $E_{zz}$ fields have a high resemblance, so do $F_{yy}$ and $E_{yy}$, $F_{yz}$ and $E_{yz}$, as can be observed in Fig. 6b. This indicates a subtle rigid body rotation.

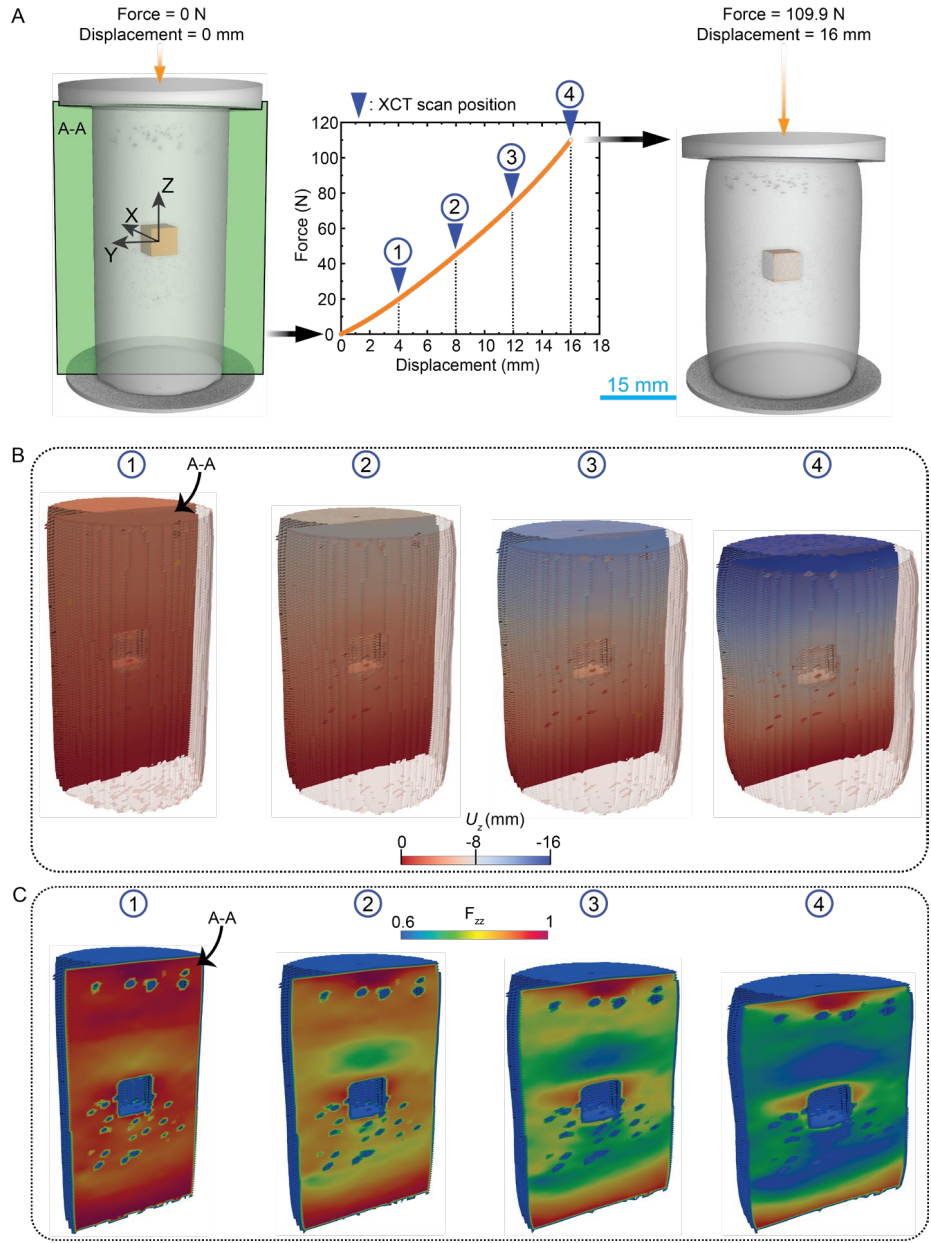

**Figure 5.** Multi-step DVC on compression test of silicone rubber cylinder with a rigid cubic acrylic inclusion. (a) XCT reconstruction of the loading setup in the initial (displacement = 0mm) and final state (displacement = 16mm). The measured force vs. displacement history is included. (b) The measured $u_z$ field at each loading step on the central plane A-A indicated in (a) and (c) the corresponding distributions of the $F_{zz}$ component of the deformation gradient.

Selected line plots are presented in Fig. 6c. The middle margin between the data points, which has a width of approximately 6 mm, is attributed to the cube inclusion. Other little gaps are due to the air voids. Along Line 1, both the $F_{yy}$ field and $F_{zz}$ field show small strain (value approaches to 1) at the boundaries because of the geometric constraint, and though unnecessary, experimental results match well with FE simulation. Similar remarks also apply to Line 2. Here, an arbitrary 3D deformation field has been quantified via a stepwise manner, and the final deformed state been elaboratively exhibited. Together with the section 2.2 and section 2.3, these three showcases progressively demonstrate the capability of this multi-step DVC methodology to quantify the full 3D deformation field in a severely deformed and nominally homogenous material.

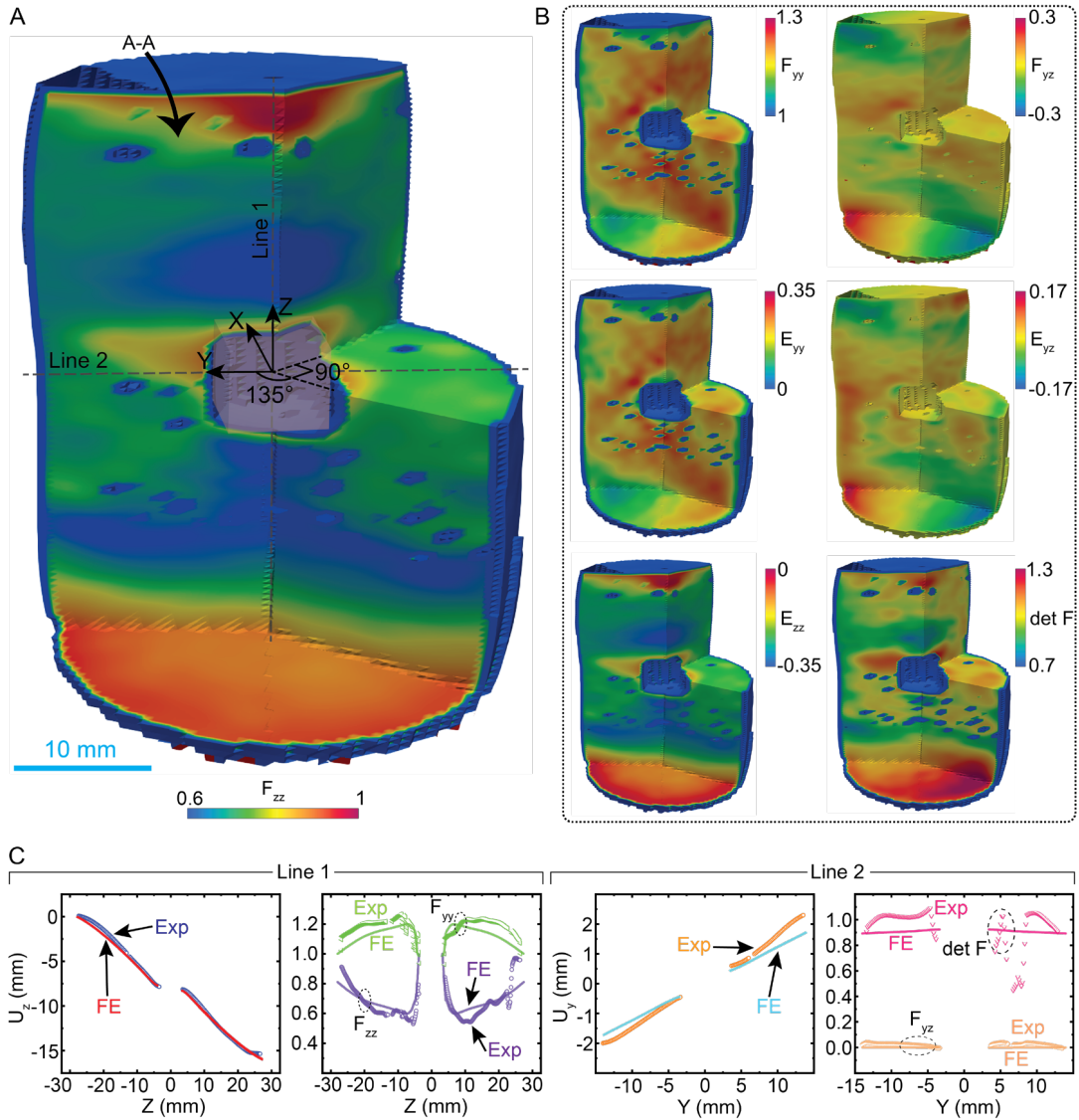

**Figure 6.** 3D deformation field in the final deformed state of the specimen with a cubic inclusion from Fig. 5 (a) The deformation gradient component $F_{zz}$ on various internal planes in the specimen. (b) Select component of the deformation gradients and Green-Lagrange strain in the view shown in (a). (c) Variations of the measured displacement deformation gradient and Green-Lagrange strain components along Lines 1 and 2 indicated in (a). FE simulation results are also included.

*2.5 A general method that applies to common engineering polymers*

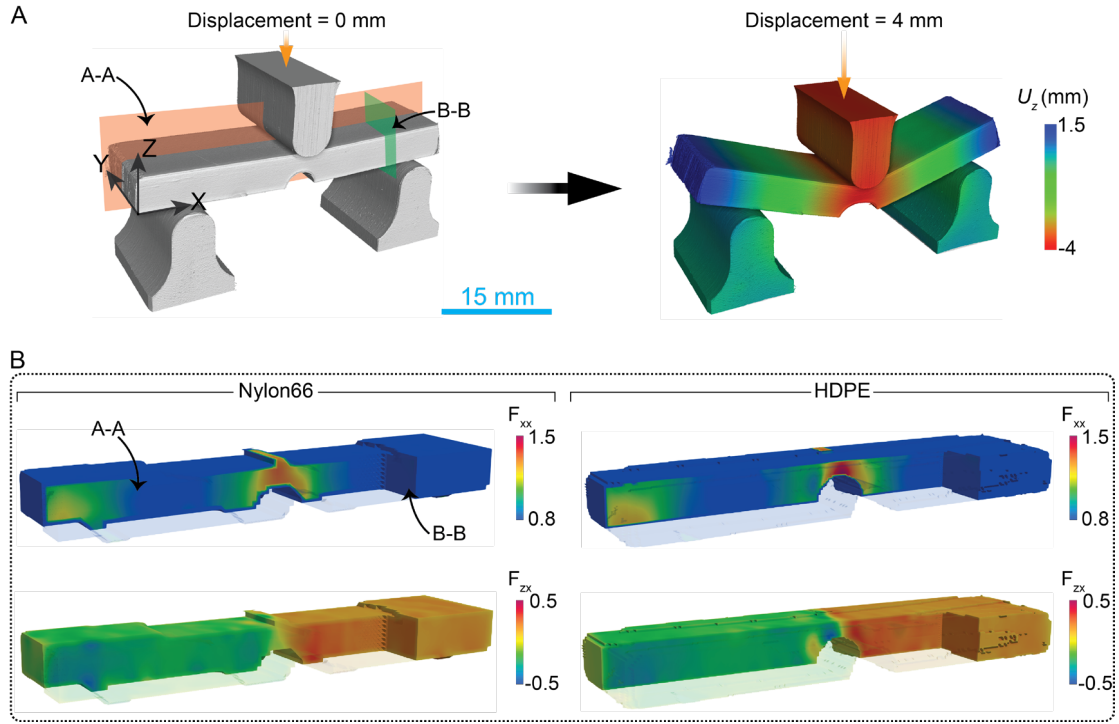

**Figure 7.** Illustration of the generality of the FETC approach for engineering polymers with measurements shown for Nylon66 and HDPE. (a) Experiment setup for 3-point bending test of a beam with a semicircular notch. The DVC inferred displacement field is shown on the deformed shape for Nylon66. (b) The measured spatial variations of selected components of the deformation gradient tensor $F$ on selected planes within the specimen. Results are shown for Nylon66 and HDPE.

In the previous sections, observe that FETC provides enhancement in volumetric grayscale contrast (Fig. 1) that can be used as tracking features for DVC analysis of deformation field in nominally homogenous materials, and the multi-step DVC methodology delivers a stepwise tracking method that can accumulate deformation field from a series of loading steps, the number of which is unlimited in theory. The first approach extends application of DVC analysis beyond materials with microstructural features (e.g. precipitation, porosity, fibrous texture, etc.) to highly homogenous materials without deliberately adding tracer particles. The second approach overcomes the limitation in the extent of deformation that can be quantified via DVC analysis. Combining these two approaches enables precise quantification of deformation field in a severely deformed and nominally homogenous material.

So far, the joint capability of the two approaches have been demonstrated using the hyperelastic silicone rubber. To prove the generality of the approach, the workflow has been tested on other nominally homogenous materials as shown in Fig. 7. Nylon66 and HDPE (High-density polyethylene) are two commonly used engineering polymers. Fig. 7a shows the loading setup, particularly, a semicircular notch through Y axis is made to create a different deformation field than that in section 2.3. Same loading setup for both the Nylon66 and HDPE. Fig. 7b shows the 3D field of $F_{xx}$ and $F_{zx}$. A high tension in X axis direction at the region around the notch can be observed for both materials. The notch in the Nylon66 beam is shallower than that in the HDPE, therefore yielding a less intense concentration. Further applications of the method can be extended to high density engineering materials like steels for understanding mechanics of metallic materials.

*Outlook*

Traditional methods to characterize constitutive models involve performing a large number of measurements under different stress-states but with specimen geometries selected such that the problem is statically determinate and spatially uniform. This implies that the stress-state is known immediately from the applied loads and the measured displacements give information of the constitutive model in a specific slice of the 3D stress space. Not only are these measurements complex (e.g., the need to sample different stress-states involves performing complex multi-axial tests, shear tests, torsion tests etc.) but also time-consuming. Since only a single stress-state is sampled in each test, many tests need to be performed to reasonably characterize a material. An alternate approach that has been proposed for several years has been to use full-field measurements on complex specimen geometries to sample a large number of stress-states in a single relatively simple uniaxial tension/compression test. Then using the full-field data the constitutive model is extracted by numerical methods that include, but not restricted to, Virtual Fields Method (VFM)[18] or more modern ML methods such as EUCLID[10]. However, these approaches have had little practical impact, as to-date the full-field data was restricted to be surface measurements using so-called 3D DIC: At most 4 of the 9 components of the deformation gradient are measurable by this method and that too only on the surface. The DIC-based full-field measurements do not give 3D stress information and the inverse problem solved by VFM, ML or other techniques is highly ill-posed. Typically, numerous *a priori* constitutive assumptions are made to regularize the inverse problem. These assumptions significantly reduce the utility of these full-field measurement methods.

Here for the first time, we have reported a laboratory-based technique that is not only capable of measuring all components of the deformation gradients within the 3D volume of specimens, but also removes the need to include artificial tracers that can change material properties and/or reduce the measurement resolution. This method, labelled FETC, combined with Eulerian/Lagrangian transformations to track very large deformations has opened the opportunity to provide the full-field data required to extract complex constitutive models from a single measurement. We have reported complex specimen geometries which sample a vast array of stress-states and give point cloud data at billions of material points in 3D. Such dataset can be utilized as input for data-driven mechanics methods with the potential to transform the fidelity and speed with which experimentally measured and validated constitutive models can be constructed.

## Acknowledgements

The authors acknowledge funding from the UKRI Frontier Research grant "Graph-based Learning and design of Advanced Mechanical Metamaterials" with award number EP/X02394X/1.

## Data Availability

The authors confirm that the data supporting the findings of this study are available within the article.

**Declaration of Competing Interest**

The authors declare that they have no competing interests in finance and personal relationships, which are related to the present work.